\documentclass{elsart}

\usepackage{amssymb}
\usepackage{amsmath}
\usepackage{amsfonts}
\usepackage{array}
\usepackage{dcolumn}
\usepackage{longtable}
\usepackage{setspace}

\usepackage{verbatim}
\usepackage{graphicx}
\usepackage[T1]{fontenc}

\newcommand{\Commentaire}[1]{}

\usepackage[english]{babel}
\usepackage{indentfirst}
\usepackage{latexsym}
\usepackage{eucal}
\usepackage{theorem}
\usepackage[cmtip,arrow]{xy}

\usepackage{pb-diagram}

\usepackage{epsfig}
\usepackage{rotate}
\usepackage[it]{subfigure}
\usepackage{color}
\usepackage{textcomp}

\usepackage{amssymb}
\usepackage{multirow}

\usepackage{times}

\newcommand{\idr}[1]{_{\text{#1}}}

\begin{document}

\begin{frontmatter}

\title{High-resolution modal analysis} 
\author{Kerem Ege\corauthref{cor1}}
\ead{kerem.ege@polytechnique.edu}
\address{Laboratory for the Mechanics of Solids, \'Ecole polytechnique, F91128 Palaiseau Cedex}

\author{Xavier Boutillon}
\ead{boutillon@lms.polytechnique.fr}
\address{Laboratory for the Mechanics of Solids, \'Ecole polytechnique, F91128 Palaiseau Cedex}
\corauth[cor1]{corresponding author}
\author{Bertrand David}
\ead{bertrand.david@enst.fr}
\address{T\'el\'ecom ParisTech / TSI - CNRS LTCI, 46 rue Barrault, 75634 Paris Cedex 13}

\begin{abstract}
Usual modal analysis techniques are based on the Fourier transform. Due to the $\Delta T.\Delta f$ limitation, they perform poorly when the modal overlap $\mu$ exceeds $30\%$. A technique based on a high-resolution analysis algorithm and an order-detection method is presented here, with the aim of filling the gap between the low- and the high-frequency domains ($30\%<\mu<100\%$). A pseudo-impulse force is applied at points of interests of a structure and the response is measured at a given point. For each pair of measurements, the impulse response of the structure is retrieved by deconvolving the pseudo-impulse force and filtering the response with the result. Following conditioning treatments, the reconstructed impulse response is analysed in different frequency-bands. In each frequency-band, the number of modes is evaluated, the frequencies and damping factors are estimated, and the complex amplitudes are finally extracted. As examples of application, the separation of the twin modes of a square plate and the partial modal analyses of aluminium plates up to a modal overlap of $70\%$ are presented. Results measured with this new method and those calculated with an improved Rayleigh method match closely.
\end{abstract}

\begin{keyword}
Modal analysis \sep Medium-frequency domain \sep Large modal overlap \sep Structural acoustics
\end{keyword}

\end{frontmatter}


\section{Introduction}
In the dynamic response of a structure, three spectral domains are usually defined: low-, mid- and high-frequency. In general, each mode is described by a modal frequency, a modal damping factor, and a modal complex amplitude distribution (see e.g \cite{EWI1984} or \cite{NOR1989}). The low-frequency domain is characterised by distinct resonance peaks and the strong modal character of the vibratory behaviour. When the frequency increases, the traditional modal identification methods cannot be used: damping increases, resonances are thus less pronounced, modes overlap and the frequency-response tends to a smooth curve. In the high-frequency domain, the vibration can be described as a \emph{diffuse wavefield} (see \emph{e.g.} \cite{SKU1958},\cite{SKU1980} or \cite{LES1988}).

\begin{figure}[ht!]
\centering
\includegraphics[width=0.7\linewidth]{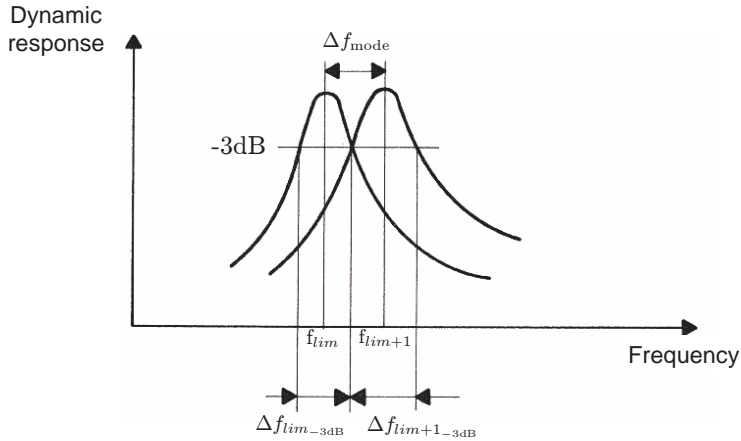}
\caption{Scheme of individual modal resonances with the same amplitude and a modal overlap factor of 100\% (after \cite{LES1988})}
\label{fig:diffwavefield}
\end{figure}
\ 

The modal overlap factor $\mu$ is the ratio between the half-power modal bandwidth and the average modal spacing: $\mu(f)=\cfrac{\Delta f_{-3\text{dB}}}{\Delta f{\idr{mode}}}$ (see \emph{e.g.} \cite{LYO1995}). The boundaries of the three spectral domains are established according to the values of $\mu$. One could define the low-frequency domain as the domain of application of modal analysis techniques: individual modes can be distinguished. It is generally admitted that the modal analysis techniques based on the Fourier transform meet their limits when the modal overlap reaches $30\%$ (see \emph{e.g.} \cite{LYO1995} or \cite{BER2003}); this is due to the $\Delta T.\Delta f$ limitation of this signal processing method.

It is commonly considered that high-frequency is reached for $\mu=100\%$ (see Fig.\ref{fig:diffwavefield}): the diffuse wavefield approximation becomes valid \cite{LES1988}. In this spectral domain, the Skudrzyk's mean-value method (\cite{SKU1980} and \cite{LAN1994}) identifies a structure by its \emph{characteristic admittance}, which is equivalent to the admittance of an infinitely extended structure. Adding other hypotheses, it is possible to apply statistical methods such as the Statistical Energy Analysis (SEA) \cite{LYO1995}, which seeks to calculate the spatial average of the response of each component of a structure by considering the equilibrium of power flows. Besides the diffuse wavefield of each subsystem, the assumptions required by SEA are that the system represents a reverberant field, that the input power sources are uncorrelated, and that the subsystems are weakly coupled (\cite{FAH1992} and \cite{LAN1992}).

In the hope of filling the gap between the low- and the high-frequency domains ($30\%\leq\mu\leq100\%$), or in effect, extending the low-frequency domain, a technique based on the high-resolution analysis algorithm ESPRIT \cite{ROY1989} and the order-detection method ESTER \cite{BAD2006} is described here. Three examples of application are presented: the separation of twin modes of a square plate (local modal overlap $\mu=200\%$) and two partial modal analyses of aluminium plates up to a modal overlap $\mu=70\%$.

In this article, modal analysis is restricted to linear systems; therefore, the impulse response $\xi(\mathbf{x},t)$ at  any point located in $\mathbf{x}$ is expected to be a sum of complex exponentials (decaying sines):
\begin{equation}
\xi(\mathbf{x},t)=\Re\left[\sum_{k=1}^{K/2}\: a_{k}(\mathbf{x}) e^{-\alpha_{k}\:t}e^{2\pi j f_{k} t +j\varphi_{k}(\mathbf{x})}\right]
\end{equation}
where $K/2$ is the number of modes, $f_k$ are the modal frequencies (in Hz), $\alpha_k$ the modal damping factors (in s$^{-1}$), $a_k(\mathbf{x})$ and $\varphi_k(\mathbf{x})$ the modal amplitudes and phases at the point of interest. 

The free dynamics of the generalised modal displacement $q_k$ is ruled by the following differential equation:
\begin{equation}
m_k\ddot{q}_k+c_k\dot{q}_k+m_k\omega_k^2q_k=0
\end{equation}
where $m_k$ is the modal mass (in kg), $c_k$ the modal damping coefficient (in kg~s$^{-1}$) and $\omega_k$ the modal angular frequency (in~rad~s$^{-1}$).

The modal damping factor $\alpha_k$ (also called modal decay constant in $s^{-1}$), the modal decay time $\tau_k$ (in s), the modal loss factor $\eta_k$ (dimensionless) and the modal damping ratio $\zeta_k$ (dimensionless) are related between them and to the above physical quantities as follows:
\begin{equation}
\alpha_k=\cfrac1{\tau_k}=\cfrac{\eta_k \omega_k}{2} \qquad \eta_k=2\zeta_k=\frac{\Delta\omega_{k,\,-3\text{dB}}}{\omega_k} \qquad \zeta_k=\cfrac{c_k}{2\:m_k\:\omega_k}
\end{equation}

If $\Delta f_{-3\text{dB}}$ is the same for two successive modes around $f$, the modal overlap $\mu$ becomes:
\begin{equation}\label{eq:mu}
\mu(f)=\frac{\eta f}{\Delta f{\idr{mode}}}=\frac1{\Delta f\idr{mode}}\,\frac{\alpha}{\pi}
\end{equation}

In practice, the modal damping factor $\alpha$ and the modal local density $\cfrac1{\Delta f{\idr{mode}}}$ are estimated in average over a narrow frequency-band centered on $f$.

Measured signals always contain some noise $\beta(t)$, which we suppose to be additive. After discretisation of $\xi(\mathbf{x},t)$ at the sampling rate $F\idr{s}=T\idr{s}^{-1}$, the signal model of the free response of the system becomes:
\begin{equation}
\xi_i(\mathbf{x})=\Re\left[\sum_{k=1}^{K/2}\: a_{k}(\mathbf{x}) e^{-\alpha_{k}\:T\idr{s}\:i}e^{2\pi j f_{k} \:T\idr{s} i+j\varphi_{k}(\mathbf{x})} +\beta_i\right]\qquad i=1 \ldots N
\end{equation}

In order to estimate the modal parameters, a high-resolution method is applied to the complex signal associated to $\xi_i(\mathbf{x})$. Historically, the Prony\cite{PRO1795} or the Pisarenko\cite{PIS1973} methods rely on the resolution of a linear prediction equation. More recent techniques assume that the signal is a sum of complex exponentials added to white noise and project the signal onto two sub-spaces. The space spanned by a finite-length vector containing successive samples is decomposed into the subspace spanned by the sinusoids (\emph{signal subspace}) and its supplementary (\emph{noise subspace}). The MUSIC\footnote{MUltiple SIgnal Classification}\cite{SCH1981}, Matrix Pencil\cite{HUA1990}, and ESPRIT\footnote{Estimation of Signal Parameters via Rotational Invariance Techniques}\cite{ROY1989} algorithms are based on this principle. The latter is chosen here since it takes into account the rotational invariance property of the signal subspace, ensuring a more precise and robust estimation.

In practice, the noise deviates from white noise and noise-whitening may prove necessary prior to analysis. A second conditioning step described by Laroche~\cite{LAR1993} consists in splitting signals into several frequency-bands: this reduces the number of (sub-)signal components to be estimated by ESPRIT within reasonable limits and is achieved by filtering the impulse response. When narrow subbands are chosen, noise-whitening usually becomes unnecessary. The next conditioning steps aim at reducing the length of each subband signal in order to keep the memory allocation low enough and the algorithm tractable in practice: each subband signal is frequency-shifted toward zero and down-sampled. The down-sampling factor is adjusted as to avoid aliasing.

In ESPRIT, the dimensions of both subspaces must be chosen \emph{a priori} and the quality of the estimation depends on a proper choice for these parameters. The best choice for the dimension of the modal subspace is the number of complex exponentials in the signal. This number is $K$, twice the number of decaying sinusoids. It is therefore advisable to estimate this number prior to the analysis. This is done by means of the recently published ESTER technique \cite{BAD2006}.

\begin{figure}[ht!]
\centering
\includegraphics[width=0.8\linewidth]{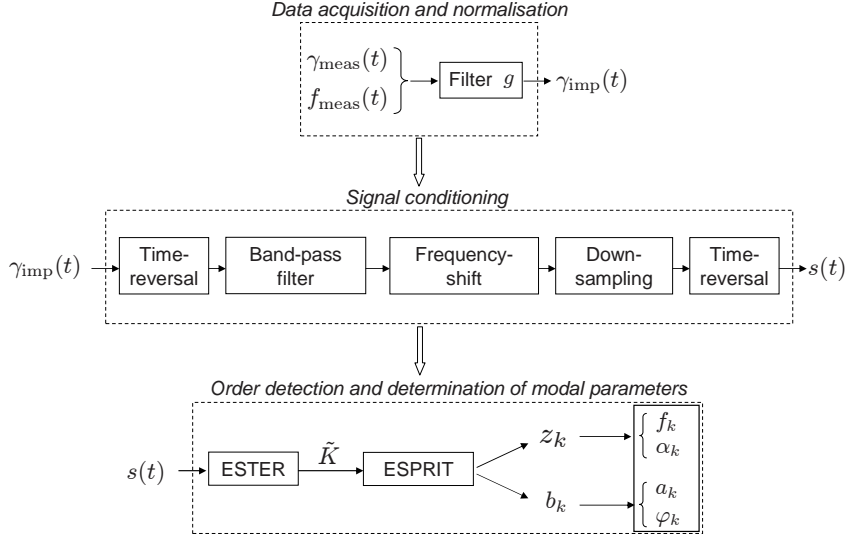}
\caption{Block diagram of the high-resolution modal analysis method.}
\label{fig:blocdiagram}
\end{figure}
\ 

The block diagram in (Fig.\ref{fig:blocdiagram}) describes the three main steps of the method:
\begin{itemize}
\item reconstruction of the acceleration impulse response (section~~\ref{sec:impulse});
\item signal conditioning (section~\ref{sec:sigcond});
\item order detection and determination of modal parameters, which constitute the heart of the method (section~\ref{sec:detmodparam}).
\end{itemize}

\section{Data acquisition and signal processing}\label{sec:datasigcond}
\subsection{Reconstruction of the acceleration impulse response}\label{sec:impulse}
A standard measuring technique in modal analysis consists in applying a pseudo-impulse force with an impact hammer on a structure and to measure both the applied force and the resulting vibration, generally by means of an accelerometer. Taking advantage of the assumed linearity of the system, the reciprocity theorem is invoked in order to obtain the modal shapes: the point of excitation is varied while the accelerometer is kept fixed, instead of the opposite. This experimental procedure has been followed throughout this article.

The analysis of free vibrations becomes a modal analysis when the response is normalised to the excitation of the system. The usual technique for this purpose is the division of the Fourier spectrum of the response by that of the excitation. In principle, the result is the Fourier transform of the impulse response of the system at the point of interest. Since our method works in the time-domain, it would be necessary to calculate the inverse Fourier transform of this response. In practice, the division of spectra proves to be dangerous for the applicability of the method: quasi-zeros in the denominator introduce high-amplitude individual components in the ratio; they may then be transformed into quasi-sinusoids by the inverse Fourier transform and appear as false modal components. In our case, the normalisation has been achieved by reconstructing the impulse response by means of an inverse-filtering technique applied to the response of the system.

The displacement $q$ of a linear mechanical system is:
	\begin{equation}
	\label{eq:ImpRep}
	q=q\idr{imp} \ast f 
	\end{equation}
where $ q\idr{imp}$ is the impulse response. The system will be considered initially at rest ($\ v(0^-)=0$) in a frame of reference such that $q(\mathbf{x},0^-)=q(\mathbf{x},0^+)=0$ at any position $\mathbf{x}$. Without loss of generality, one may also consider $q\idr{imp}(0^+)=0$. It should be noted that $v\idr{imp}(0^+)$ and $v(0^+)$ are not zero in general.

Denoting Laplace transforms by uppercase letters, the generic expression $\mathcal{L} \left[\dfrac{df}{dt}\right]=u\mathcal{L}(f)-f(0^+)$ of the Laplace transform of the time-derivative of a function yields:
\begin{equation}
V = u Q\idr{imp} \cdot F =\ \mathcal{L}(v\idr{imp})\cdot \mathcal{L}(f)\qquad\Rightarrow \qquad v=v\idr{imp} \ast f\nonumber
\end{equation}

The impulse acceleration response is given by:
\begin{align}
\Gamma &=u V\idr{imp}\cdot F - v(0^+)\nonumber\\
       &=\left[\mathcal{L}(\gamma\idr{imp})+v\idr{imp}(0^+)\right]\cdot F - v(0^+)\nonumber\\
       &= \mathcal{L}(\gamma\idr{imp})\cdot \mathcal{L}(f)+v\idr{imp}(0^+) \mathcal{L}(f) - v(0^+)\nonumber\\
\Rightarrow\quad \gamma&=\left\{\gamma\idr{imp}\ast f\right\} + v\idr{imp}(0^+) \cdot f - v(0^+) \cdot\delta\label{eq:AccMeas}
\end{align}

\begin{figure}[ht!]
\centering
\includegraphics[width=0.85\linewidth]{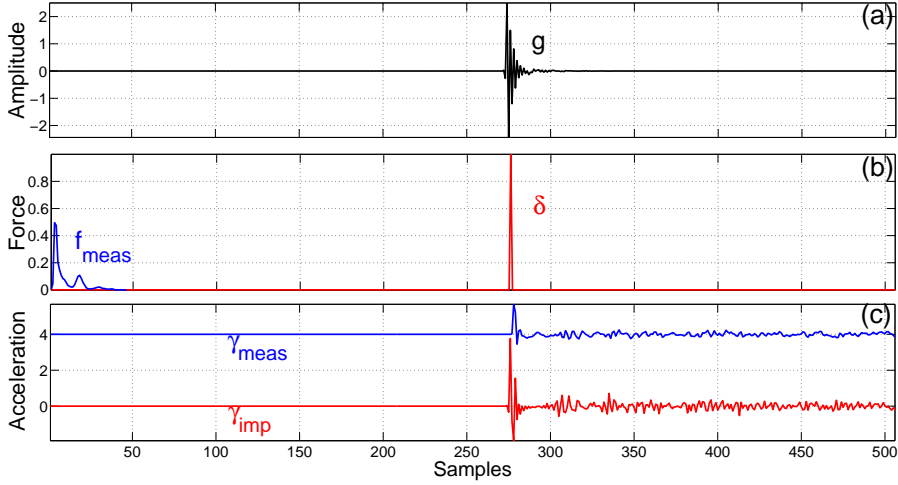}
\caption{Reconstruction of one impulse acceleration response (arbitrary units). (a) Optimised filter $g$, with $p=276$. (b) Measured force {\color[rgb]{0,0,1}$f\idr{meas}$} ($m=46$) and reconstructed pulse {\color[rgb]{1,0,0}$\delta$} shifted by $p$ samples. (c) Measured response {\color[rgb]{0,0,1}$\gamma\idr{meas}$} and reconstructed response {\color[rgb]{1,0,0}$\gamma\idr{imp}$}. }
\label{fig:Impulse}
\end{figure}

Given the measurements of the force $f\idr{meas}$ and the acceleration $\gamma\idr{meas}$, the impulse response $\gamma\idr{imp}$ is estimated as follows. The first step consists in finding a finite-impulse-response (FIR) filter $g$ that transforms the force signal $f\idr{meas}$ into a normalised pulse (Fig.~\ref{fig:Impulse}):
\begin{equation}
\label{eq:g}
f\idr{meas}\ast g=\delta_{\stackrel{p}{\rightarrow}}
\end{equation}

Here, $g$ stands for the impulse response of the filter in the continuous time-domain and $\delta_{\stackrel{p}{\rightarrow}}$ represents the Dirac impulse shifted in time for causality reasons: $\delta_{\stackrel{p}{\rightarrow}}\:=\:\delta(t-p/F\idr{s})$.

\ 

When a hammer is used to excite the system, the excitation duration is finite and an upper bound for the number $m$ of samples in $f{\idr{meas}}$ can be given with certainty. After discretisation, the convolution equation Eq.~\eqref{eq:g} defines a system of linear equations. The best solution, for example in the least mean-square sense, can be found by commonly available algorithms. We have chosen a filter with $11\,m+1$ coefficients and $p=6\,m$.

In a second step, Eq.~\eqref{eq:AccMeas} is applied to the measured values of the acceleration $\gamma=\gamma\idr{meas}(t)$ and the force $f=f\idr{meas}(t)$. After convolution of Eq.~\eqref{eq:AccMeas} by $g$ and substitution of $f\idr{meas}(t)\ast g$ by $\delta_{\stackrel{p}{\rightarrow}}$, the result is shifted up in time by $p T_s$. One obtains an estimation of the impulse acceleration response $\gamma\idr{imp}$:
\begin{equation}
\gamma\idr{imp}=\left\{\gamma\idr{meas}\ast g\right\}_{\stackrel{-p}{\rightarrow}}\:-\: v\idr{imp}(0^+)\cdot\delta+v(0^+)\cdot g_{\stackrel{-p}{\rightarrow}}
\end{equation}

When the system is excited by a continuous force (no shock), $v(0^+)$ is $0$ and the above expression becomes simpler. Otherwise, $v(0^+)$ can be estimated by integrating $\gamma(t)$. In practice, it may be difficult to extract the signal from the noise in $\gamma\idr{meas}$ and obtaining a precise value of $v(0^+)$ may turn difficult. The solution consists in defining the origin of time slightly before the impact hammer touches the structure (this is generally obvious by inspection); this guarantees that $\gamma(T_s)$ and $v(0^+)$ are truly $0$.

The process of retrieving the acceleration impulse response is illustrated in Fig.~\ref{fig:Impulse}. The first sample of the impulse response cannot be retrieved since $v\idr{imp}(0^+)$ is not known. If necessary, it could be reconstituted at the end of the modal analysis and the corresponding correction be applied to the modal amplitudes and phases.

\subsection{Signal conditioning}\label{sec:sigcond}
\subsubsection{Reduction of the number of points}
The number of operations in the ESPRIT algorithm is $O(N\,^3)$ and the computing duration is excessively long for a large number $N$ of samples. Numerical instabilities may also appear. In order to overcome these problems, we adopt the procedure proposed by Laroche \cite{LAR1993} and reviewed in the introduction: band-filtering, frequency-shifting, and decimating. A few minor transformations are introduced.

It is advisable to evaluate roughly the spectral density of modes\footnote{This may be done by mechanical reasoning or by extrapolating the low-frequency analysis, for example.}. This helps to define frequency-bands that contain less than say $\tilde{K}=25$ complex components \cite{BAD2005}. A band-pass filter between $f\idr{l}$ and $f\idr{h}$ is designed for each band. Although not as efficient as IIR filters, FIR filters are preferred because their transfer function has no pole and therefore, does not introduce spurious modes into the signal. Various techniques for synthesising the filter are available. We have chosen the Blackman window.

The signal is then filtered as follows. An impulse response encounters a large variation at $t=0$ and decreases afterwards. In order to minimise the effect of the transient response of the filter, the signal is time-reversed prior to convolution with the FIR $h$ of the filter. This does not alter its spectrum. Convolution adds a number of samples equal to the length of $h$, at the end of the reversed signal. These points must be removed from the \emph{beginning} of the signal after it is time-reversed again (see below). Once filtered, only $\tilde{K}/2$ modes are kept. However this number is still to be determined with precision. The amplitudes and phases of the modes at the measured point are altered by the filtering and their transformed values are written $\tilde{a}_k$ and $\tilde{\varphi}_k$.

The Hilbert transform of the filtered signal is computed in order to eliminate the negative-frequency content of the spectrum which would cause aliasing problems in the next steps of the procedure. We have used the \verb hilbert  function proposed by Matlab$^\circledR$. The procedure does not include any spectrum division; the Gibbs phenomenon (very rapid oscillations) associated to the Fourier truncation done in this procedure is limited to the very beginning and to the very end of the transformed signal. Because of a very fast decay rate, it never proved problematic in practice (in other words: no pseudo-poles were added by the Fourier truncation). The signal now contains $\tilde{K}/2$ complex exponentials whose frequencies are between $f\idr{l}$ and $f\idr{h}$.

This signal is multiplied by $\exp(-2\pi j \:f\idr{d}\: i\: T\idr{s})$, with $i=1, \ldots, N$. This operation shifts the spectrum by $f\idr{d}$ which is chosen slightly less than $f\idr{l}$. The spectrum of the result is now limited by $f'\idr{l}=f\idr{l}-f\idr{d}$ and  $f'\idr{h}=f\idr{h}-f\idr{d}$. As a matter of preference, we have then taken the real part of this complex signal. This produces a symmetrical spectrum with $\tilde{K}$ components between $-f'\idr{h}$ and $+f'\idr{h}$.

According to the sampling theorem, the signal may now be down-sampled at a sampling frequency lower than $F\idr{s}$, reducing the number of points to analyse. In principle, the decimating factor $d$ could be chosen up to $F\idr{s}/2f'\idr{h}$; in practice, a safety margin is kept and the decimating factor that we have used was approximately $F\idr{s}/6f'\idr{h}$. Requirements on the minimum number of points in the signal add other constraints on the decimating factor (see below).

After decimation, time-reversing, and the removal of extra points (see above), the signal takes the form:
\begin{equation}
s_i=x_i+\beta_i=\sum_{k=1}^{\tilde{K}}\: b_k\:z_k^i  +\beta_i \qquad i=1\  \ldots\ \tilde{N}=\frac{N}{d}
\end{equation}
where $x_i$ is the modal signal (to be determined), $z_k=e^{-\tilde{\alpha}_k\,T\idr{s}\, d +2\pi j \tilde{f}_k \,T\idr{s}\,d}$ are its so-called poles ($\tilde{f}_k=f_{k}-f\idr{d}, \tilde{\alpha}_k=\alpha_{k}$), $b_k=\tilde{a}_k e^{j\tilde{\varphi}_k}$ are the complex amplitudes, and $\tilde{K}$ is the number of complex exponentials to be found. ESPRIT requires that the number $\tilde{N}$ of signal points be more than $2\tilde{K}$.

\subsubsection{Noise whitening}
In principle, the results of the ESPRIT analysis correspond to the complex frequencies of the signal only if the additive noise $\beta$ is white. In practical cases, the noise is white to first order in any narrow band, hence the interest of subband filtering presented above. For wide frequency-bands, including a noise-whitening step in the signal-conditioning procedure may improve the precision of the modal results. A method proposed by Badeau\cite{BAD2005} consists in estimating the power spectral density of the noise for each frequency-band and to deduce from it the corresponding whitening filter. The Fourier spectrum is computed first and a \emph{rank filter}\footnote{In a rank filter, the data are sorted by ascending orders. The output value is the $r^\text{th}$ lowest data value, where $r$ is the rank order of the filter.} is used in order to smooth the spectrum. Then, the estimator of the autocovariance function is found by calculating the inverse Fourier transform of this filtered spectrum. A linear prediction on this estimator gives the coefficients of the whitening filter that can, finally, be applied to the original signal.

This noise-whitening treatment did not prove necessary in the applications presented here.

\subsection{Determination of modal parameters}\label{sec:detmodparam}
\subsubsection{Order detection}
\label{sec:Ester}
\label{sec:OrderDetection}
As mentioned above, the best choice for the dimension of the modal subspace to be given to the ESPRIT algorithm is $\tilde{K}$. Obviously, a larger value may also be chosen: some of the effective noise will be partly projected on to modal subspace, producing very weak or highly attenuated components. A choice smaller than $\tilde{K}$ for the dimension of the modal subspace would introduce errors in the estimation of the modal components.

In order to estimate the number of complex exponentials (that is: twice the number of modes) in the signal, we have used the ESTER (ESTimation ERror) procedure by Badeau \cite{BAD2006} which is schematically presented here. One notes that the first steps of this procedure are common with those of the ESPRIT algorithm \cite{ROY1989}.

The $\tilde{N}$ signal data $s_i$ and the modal signal samples $x_i$ are written in the form of Hankel matrices:
\begin{equation}
S=\left(\begin{array}{cccc}s_1 &s_2 &\dots &s_l\\
\vdots&&&\vdots\\
s_n &s_{n+1} &\dots &s_{\tilde{N}} \\\end{array}\right)
\qquad
X=\left(\begin{array}{cccc}x_1 &x_2 &\dots &x_l\\
\vdots&&&\vdots\\
x_n &x_{n+1} &\dots &x_{\tilde{N}} \\\end{array}\right)
\end{equation}
with $l=\tilde{N}-n+1$, $n$ being the sum of the dimensions of the signal and noise subspaces.

It has been shown (see \emph{e.g.} \cite{LAR1993} or \cite{BAD2005}) that:
\begin{itemize}
\item the estimation is optimal when $n=\tilde{N}/3$ or $n=2\tilde{N}/3$,
\item the estimation quality is rapidly degrading outside this interval, 
\item the estimation is only slightly degraded for $n\in[\tilde{N}/3,2\tilde{N}/3]$.
\end{itemize}
In consequence, we have systematically chosen $n=\tilde{N}/2$.

The correlation matrices are formed (computed in the case of $R\idr{ss}$):
\begin{equation}
R\idr{ss}= \dfrac{1}{l}\:S\:S^H \qquad R\idr{xx}= \dfrac{1}{l}\:X\:X^H
\end{equation}

For additive white noise with variance $\sigma^2$:
\begin{equation}
\mathbb{E}\left[R\idr{ss}\right]= R\idr{xx} +\sigma^2 \ I
\label{eq:SumValProp}
\end{equation}
which shows that the eigenvectors of $R\idr{xx}$ are among those of $R\idr{ss}$ in the limit of perfect estimation.

The algorithm ESPRIT needs the $\tilde{K}$ eigenvectors of $R\idr{xx}$ to determine the poles \{$z_k$\}.
It is now shown how to find both $\tilde{K}$ and those eigenvectors.

The eigenvalues $\lambda_m$ ($m=1, \ldots, n$) and the corresponding eigenvectors\linebreak $\{w_1,\ldots, w_n\}$ of $R\idr{ss}$ are computed. It can be shown \cite{ROY1989} that
\begin{itemize}
\item the eigenvalues are real and positive,
\item eigenvalues associated with the noise subspace are equal to $\sigma^2$ (nearly equal for a non-white noise).
\end{itemize}

Ordering the eigenvalues in decreasing order naturally selects the ones associated with the modal signal: in principle, $\tilde{K}$ is the number of eigenvalues that verify $\lambda_{m}>\sigma^2$ (see Eq.~\ref{eq:SumValProp}). The ESTER criterion presented below is more robust than this condition for the determination of $\tilde{K}$.

$W(p)$ is defined as the matrix formed by columns $w_k$: $\{w_1,\ldots, w_p\}$ with $p<n$. The matrix $W_{\uparrow}(p)$ is defined by removing the first line of $W(p)$ and $W_{\downarrow}(p)$ is defined by removing the last line of $W(p)$. The following matrix $\Phi$ and quantity $E(p)$ are formed:
\begin{align}
\Phi(p)= W_{\downarrow}(p)^\dag \ W_{\uparrow}(p) \quad
E(p)=W_{\uparrow}(p) - W_{\downarrow}(p) \Phi(p)
\end{align}
where $W^{\dag}$ is the pseudo-inverse of $W$. 

The ESTER criterion defines $\tilde{K}$ as the highest $p$ maximising $J(p)=1/||E(p)||^2$. In other words, $\tilde{K}$ is found as the highest number such that $E(\tilde{K})$ approaches 0, which corresponds to the so-called rotation invariance of $W$.

\begin{figure}[ht!]
\centering
\includegraphics[width=0.65\linewidth]{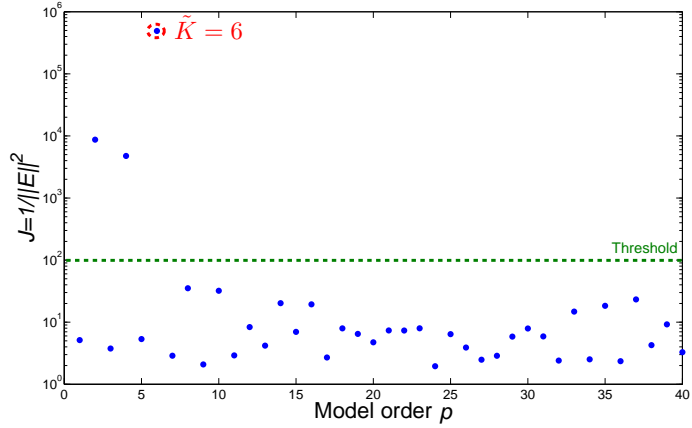}
\caption{Application of the ESTER criterion to a signal made of three damped sinusoids and additive noise ($\text{SNR}=50$~dB, see Table~\ref{tab:EspritTest} for the other parameters). The detection threshold for ESTER is chosen to 100. The value ${\tilde{K}}=6$ (corresponding to 3 modes) is clearly detected.}
\label{fig:Ester}
\end{figure}

The case of a synthesised signal with 3 sinusoids and added noise (signal to noise ratio $SNR=50$ dB) is shown in Fig.~\ref{fig:Ester} (see Table~\ref{tab:EspritTest} for the modal components parameters). A threshold $J_t$ is chosen (here: $10^2$), in correspondence with $SNR$ and $\tilde{K}$ is considered as the highest value of $p$ for which $J(p)>J_t$ (here: $p=6$). This criterion proves to be very robust.
\ 

\begin{table}[ht!]
\centering
\begin{tabular}{l c c c c c c}
   \hline
   
   \hline
    \ & \multicolumn{3}{c}{Parameters of the test signal} &\multicolumn{3}{c}{Parameters estimated by ESPRIT}\\
 \hline
 $f_k~[\text{Hz}]$ & 2078.10 & 2082.30 & 2087.10 & 2078.11 & 2082.31 & 2087.12\\
 $\alpha_k~[\text{s}^{-1}]$ & 28.00 & 31.00 & 27.00& 27.96 & 30.72 & 27.02 \\
 $a_k$ & 1.00 & 0.80 & 0.40& 1.00 & $0.77 $ & 0.40\\
 $\varphi_k~[\text{rad}]$& $\cfrac{\pi}{2}\ (\approx1.57)$ & $-\cfrac{\pi}{3}\ (\approx-1.05)$ & $-\cfrac{\pi}{6}\ (\approx-0.52)$& 1.56& -1.05 & $-0.54$\\
 \hline
 
 \hline
 \end{tabular}\caption{Comparison between true and estimated parameters of a synthetic signal.}
\label{tab:EspritTest}
\end{table}
\

\subsubsection{Modal frequencies, modal damping factors, and complex amplitudes}
Once the number of modes $\tilde{K}/2$ has been estimated, the $\tilde{K}$ first columns of $W(n)$ are extracted to form $W=W(\tilde{K})$, the matrix of the eigenvectors of $R_{xx}$. The purpose of the ESPRIT procedure is to derive the so-called poles ${z_k}$ from this information on the modal signal. The main steps are schematically recalled here (for a demonstration, see \cite{ROY1989}):
\begin{itemize} 
\item The Vandermonde matrix $V^n$ and the diagonal matrix $D$ are formed with the ${z_k}$: 
$$V^n=\left(\begin{array}{cccc}1 &1 &\dots &1\\
z_1 &z_2 &\dots &z_{\tilde{K}}\\
z_1^2 &z_2^2 &\dots &z_{\tilde{K}}^2\\
\vdots&&&\vdots\\
z_1^{n-1} &z_2 ^{n-1} &\dots &z_{\tilde{K}} ^{n-1} \\\end{array}\right) \qquad
D=\left(\begin{array}{ccc} z_1 &&(0) \\
&\ddots&\\
(0)	&&z_{\tilde{K}}\end{array}\right)$$

Their rank is ${\tilde{K}}$ and they verify: 
\begin{equation}\label{eq:RotInv}
V^n_{\uparrow}=V^n_{\downarrow} D
\end{equation}
where the matrice $V^n_{\uparrow}$ (repectively $V^n_{\downarrow}$) are formed by eliminating the first row (respectively the last row) of $V^n$.

\item The rank of $W$ is also ${\tilde{K}}$ and therefore, a base-change matrix $C$ can be defined by:
\begin{equation}
\label{eq:BaseChange}
V^n=W\ C
\end{equation}

Shifting this equation up and down yields $V^n_{\downarrow}=W_{\downarrow}\ C$ and $W_{\uparrow}=V^n_{\uparrow}\ C^{-1}$ 

\item Using Eq.~\ref{eq:RotInv} yields: 
\begin{equation}
W_{\uparrow}=W_{\downarrow} C\ D\ C^{-1}
\qquad \Rightarrow \qquad  W_{\downarrow}^\dag \ W_{\uparrow} = C\ D\ C^{-1}
\label{eq:RotInvW}
\end{equation}

This equation, denoting a so-called rotation-invariance property of $W$, shows that the poles ${z_k}$ are the eigenvalues of $W_{\downarrow}^\dag\ W_{\uparrow}$.
\end{itemize}
The frequencies and damping factors of the response signal are:
\begin{align}
f_{k}=\frac{\arg(z_k)}{2\pi}\:\frac{F\idr{s}}{d}+f\idr{d} \qquad
\alpha_{k}=-\frac{F\idr{s}}{d}\:\ln|z_k| 
\end{align}

The final step consists in the determination of the amplitudes and phases of the modal components. To this end, the $\tilde{N}\times {\tilde{K}}$ Vandermonde matrix $V^{\tilde{N}}$ is formed. The complex amplitudes $b_k$ are the best solution, in the  least-mean-square sense, of the equation:
\begin{equation}
V^{\tilde{N}}\ \left[\begin{array}{c}b_1\\ \vdots\\b_{\tilde{K}}\end{array}\right]=\left[\begin{array}{c}
s_1\\ \vdots \\s_{\tilde{N}}\end{array}\right]
\end{equation}

The amplitudes and phases of the response are:
\begin{eqnarray}
a_{k} &&=\frac{|b_k|}{|H(f_k)|}\\
\varphi_{k} &&=\arg(b_k)-\arg[H(f_k)] 
\end{eqnarray}

Table~\ref{tab:EspritTest} shows the estimated results for the synthetic signal described above. The error is generally less than 1\% (4\% for the phase of the third component and amplitude of the second component).

\section{Applications}
Partial modal analyses are shown in three cases:
\begin{itemize}
\item a square aluminium plate (A) with localised damping: twin modes with $\mu\simeq200\%$,
\item a rectangular aluminium plate (B) in the mid-frequency domain ($30\%\leq\mu\leq50\%$),
\item a rectangular aluminium plate (C) in the mid-frequency domain ($45\%\leq\mu\leq70\%$).
\end{itemize}
\subsection{Experiments}
A pseudo-impulse force is applied by means of an impact hammer (\emph{P.C.B. Piezotronics 086D80}). The acceleration is measured with an accelerometer (\emph{Brüel \& Kj\ae r - ENDEVCO, Isotron 2250A-10}). In all cases, boundary conditions are kept as close as possible to "free-free". The point of excitation is varied (see section~\ref{sec:impulse}) whereas the vibration measurements are made at a single point, in the vicinity of a corner of the plate. Under the chosen boundary conditions, this location is not on any of the nodal lines.

The signal analysis described in the previous sections is applied independently to each pair of measurements \{$f\idr{meas},\gamma\idr{meas}$\}. The frequency and the damping factor of each mode is taken as the weighted mean of all the estimated values. Weights are the estimations of the amplitude at each point: this gives less importance to the less precise estimations in the nodal regions.

The masses of the plates (A), (B), and (C) are respectively 0.48~kg, 5.5~kg, and 22.5~kg. Despite its relatively low mass (0.4~grams), the accelerometer causes a slightly negative shift of the modal frequencies. This phenomenon was evaluated quantitatively on plate (A) by placing a second accelerometer with the same mass just opposite to the first one. A frequency drift of  $-\,0.7$~Hz was observed for the (2,1)-mode and of $-\,0.3$~Hz for the (1,2)-mode, both at approximately 180~Hz (see section~\ref{sec:plateA}). To first order, the mass loading effect of one accelerometer can be corrected by adding the measured drift to the modal frequencies measured in the situation with one accelerometer only. For plate (B) (5.5~kg), a negative drift of less than 0.1~Hz was observed for the modes of interest, around 600~Hz. For the heaviest plate (C) (22.5~kg) a negative drift of about 0.1~Hz was observed around 1600~Hz, close to the uncertainty of our method for these high frequencies (see section~\ref{sec:PlateC}).

\subsection{Theoretical modal determination}
Only approximate solutions are known for the frequencies and the shapes of the conservative modes of a thin isotropic rectangular plate with free-free boundary conditions. Warburton~\cite{WAR1954} combined a Rayleigh method with characteristic beam functions to obtain a simple approximate expression. In this approach, plate modes are assumed to be the product of beam functions:
\begin{equation}\label{eq:rayleigh}
W_{(m,n)}(x,y)=X_m(x)Y_n(y)
\end{equation}
where $x$ (resp. $y$) corresponds to the shorter length (resp. longer) of the plate and $X_m$ (resp. $Y_n$) is the $m$ (resp. $n$)-th normal mode of a beam with the same boundary conditions as the corresponding edges of the plate. The frequency accuracy is excellent for plates with constrained edges but it is less so when one or more edges are left free. Kim \& Dickinson \cite{KIM1985} provide an improved approximate expression by using the Rayleigh method in connection with the minimum potential energy theorem: the deflection $W_{m,n}(x,y)$ includes three terms (see~\ref{sec:Appendix_beam}). For comparison with experiments, we have retained this method since the errors on modal frequencies are known to be less than 1\% \cite{HUR2007} with tractable frequency expressions. In our experiments, the uncertainties and approximations are such (see below) that more precise methods (Rayleigh-Ritz method, superposition, exact series solutions, finite element analysis, see Hurlebaus~\cite{HUR2007} for an exhaustive comparison) are not necessary.

\subsection{Separation of the twin modes of a square plate (A): low-frequency and high or low modal overlap.}\label{sec:plateA}
\begin{figure}[ht!]
\centering
\includegraphics[width=0.5\linewidth]{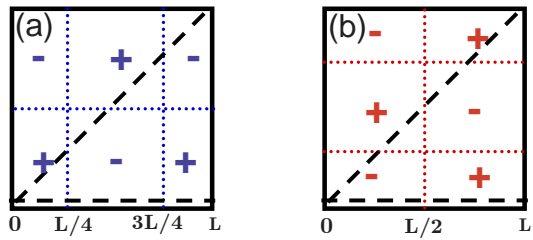}
\caption{Twin modes of a square plate ($\text{L}=300$~mm): (a)~(2,1)-mode; (b)~(1,2)-mode. Dotted lines: nodal lines. Dashed lines denote where modal analyses are performed.}
\label{fig:mode2_1}
\end{figure}

An aluminium square plate (\emph{AU4G}, $300 \times 300 \times 1.9$ mm$^3$) is suspended by rubber bands. A block of foam is glued in the centre in order to increase damping. In principle,  modes $(2,1)$ and $(1,2)$ have the same modal frequency (twin modes), and their modal shapes are similar under a $90^\circ$ rotation (Fig.~\ref{fig:mode2_1}). In practice, modal frequencies and dampings are slightly different due to imperfections in symmetry and isotropy. Here, the modal frequencies of the two modes are $\approx178$~Hz and $\approx181$~Hz, corresponding to a local (apparent) modal density of $\approx3$~modes~Hz$^{-1}$.

\begin{figure}[ht!]
\centering
\includegraphics[width=1\linewidth]{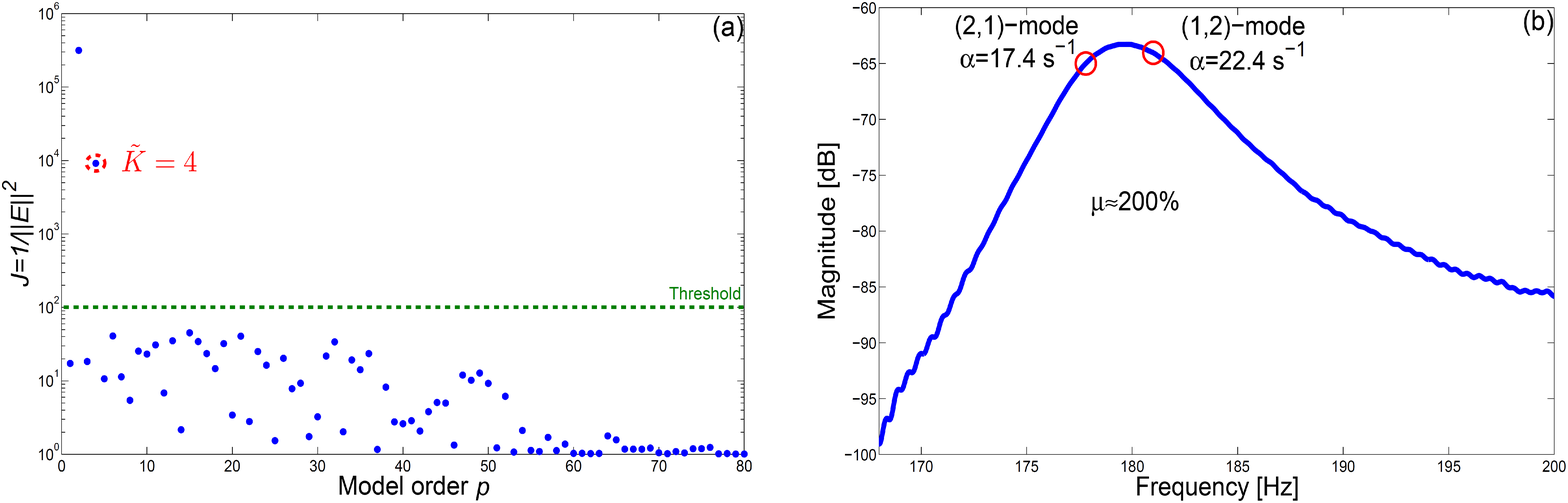}
\caption[ESTER criterion, plate (A)]{Analysis of the first twin modes of square plate (A). (a)~ESTER criterion on the response signal in point 9: the value ${\tilde{K}}=4$ (two modes) is detected. (b)~Amplitude of the Fourier spectrum at the same point. The length of the "useful" signal is $\approx 2$~s (it is masked by noise afterwards) and increasing amounts of zero-padding were tried: beyond a total length of 20~s for the analysis window (as retained here), the spectrum does not change appreciably. \textcolor[rgb]{1,0,0}{$\circ$}~marks: modes estimated by ESPRIT.} 
\label{fig:analysis_plateA}
\end{figure}

\ 
The analysis is done along one side ($y=0$) and along one diagonal as shown in Fig.~\ref{fig:mode2_1}. Plate vibrations are damped by means of a block of foam glued in the centre. The modal damping factors $\alpha$ are $\approx20$~s$^{-1}$, corresponding to a modal overlap of $\approx200$\%. The ESTER procedure reveals two modes in the 170-200~Hz frequency-band, as shown in Fig.~\ref{fig:analysis_plateA}(a). They are undistinguishable in a typical Fourier spectrum (Fig.~\ref{fig:analysis_plateA}(b)). The estimations of the modal parameters are given in Table~\ref{tab:ModalComp_plateA}.
 
\begin{table}[ht!]
\centering
\begin{tabular}{l c c c}
   \hline
   
   \hline
\ &\ &(2,1)-mode&(1,2)-mode\\
	\hline
\multirow{2}{28mm}{Plate with extra damping}& $f$ (Hz)&177.8 & 181.0\\
\ & $\alpha$ (s\textsuperscript{-1})&17.4 & 22.4\\
 \hline
\multirow{2}{28mm}{Plate without extra damping}& $f$ (Hz)& 178.1&181.4\\
\ & $\alpha$ (s\textsuperscript{-1})& 2.6 &3.9\\
 \hline
 
 \hline
 \end{tabular}
\caption{Plate (A), with and without artificial extra damping: estimations of the modal parameters of the twin modes $(2,1)$ and $(1,2)$.}\label{tab:ModalComp_plateA}
\end{table}

\begin{figure}[ht!]
\centering
\includegraphics[width=0.6\linewidth]{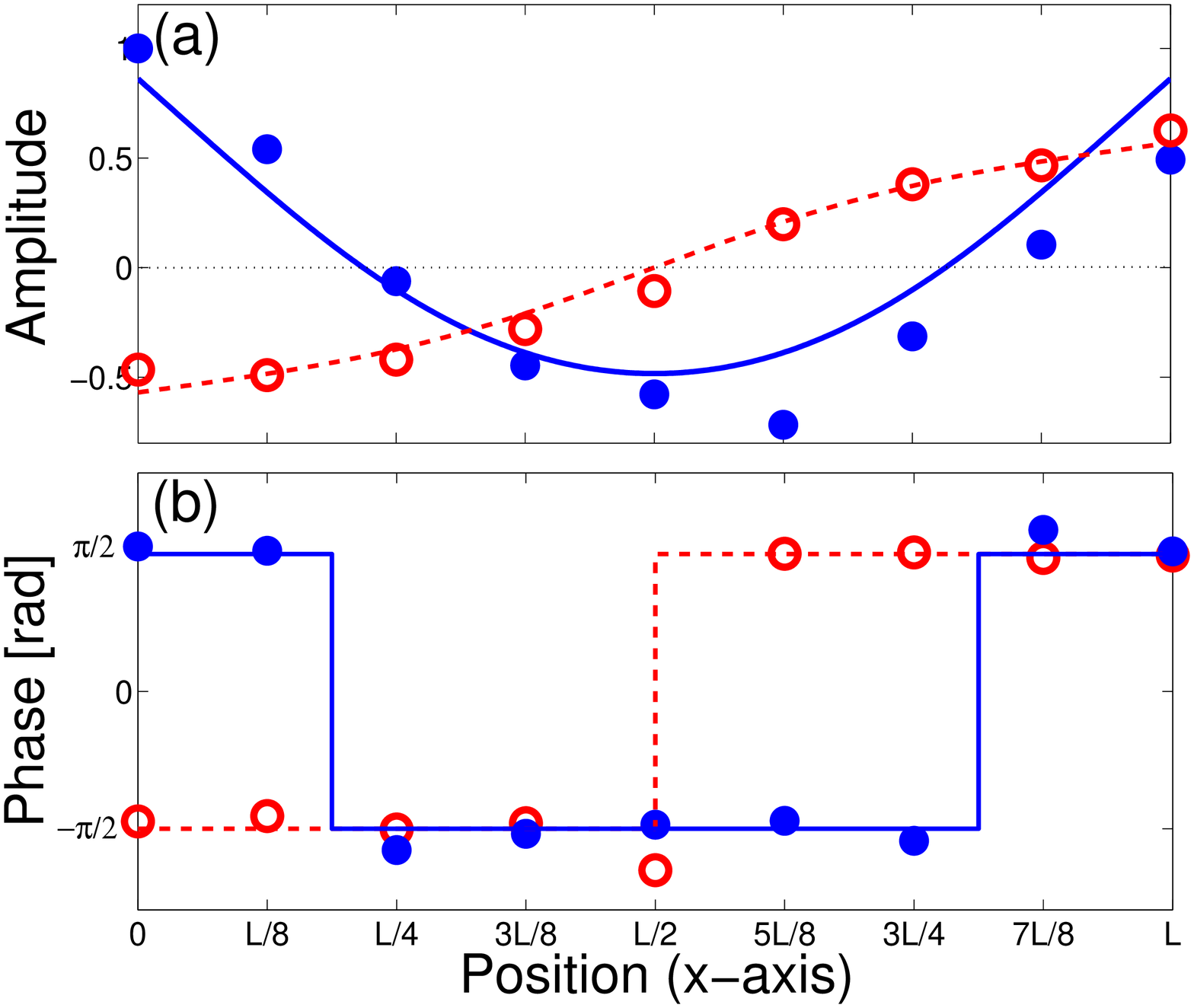}
\caption[Separation of twin modes along one side]{Separation of the twin modes along one side of square plate (A). (a)~Normalised "signed" amplitudes; (b)~Phase. \textcolor[rgb]{0,0,1}{$\bullet$}~marks: \emph{measured} (2,1)-mode. \textcolor[rgb]{0,0,1}{Solid}~line: \emph{theoretical} conservative (2,1)-mode. \textcolor[rgb]{1,0,0}{$\circ$}~marks: \emph{measured} (1,2)-mode. \textcolor[rgb]{1,0,0}{Dashed}~line: \emph{theoretical} conservative (1,2)-mode.}
\label{fig:defmod_2_1_delta1}
\end{figure}

\begin{figure}[ht!]
\centering
\includegraphics[width=0.6\linewidth]{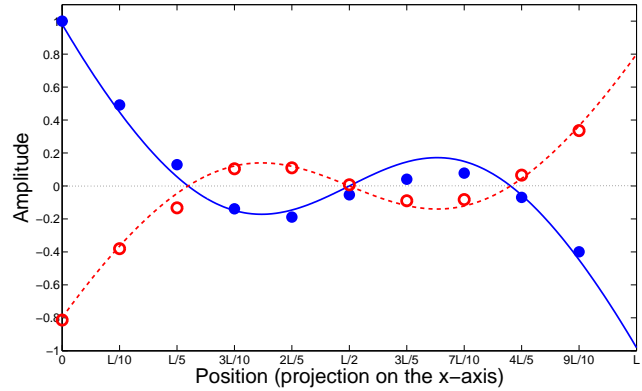}
\caption[Separation of twin modes along one diagonal]{Separation of the twin modes along one diagonal of plate: normalised "signed" amplitudes. \textcolor[rgb]{0,0,1}{$\bullet$}~marks: \emph{measured} (2,1)-mode. \textcolor[rgb]{0,0,1}{Solid}~line: \emph{theoretical} conservative (2,1)-mode. \textcolor[rgb]{1,0,0}{$\circ$}~marks: \emph{measured} (1,2)-mode. \textcolor[rgb]{1,0,0}{Dashed}~line: \emph{theoretical} conservative (1,2)-mode.}
\label{fig:diag2_1}
\end{figure}

With the sign of the modal phase attributed to the amplitude, the modal "signed amplitudes" along one side are displayed in Fig.~(\ref{fig:defmod_2_1_delta1}-a) together with the theoretical modal amplitudes for the conservative case (dashed line). Here and in what follows, the measured modal shapes are normalised to a maximum of $1$. The amplitudes of theoretical modal shapes are adjusted to yield a best fit (in the least-mean-square sense) to the experimental data. The modal phases are given in Fig.~(\ref{fig:defmod_2_1_delta1}-b). The modes can be considered as clearly and adequately separated in this case of very high local modal overlap. The differences between measured and theoretical amplitude curves of the (2,1)-mode (particularly noticeable for $x>L/2$) are due to the mass of the accelerometer placed at $x=y=0$. The light mass (0.4~grams) slightly modifies the modal shapes. We observed that adding one similar accelerometer at $(x=L,y=0)$ removes the asymmetry of the measured modal shape.

\  
\ 
Without the block of foam, damping factors are around 3~s$^{-1}$, corresponding to an overlap of $\approx30\%$. The estimations of modal parameters are given in Table~\ref{tab:ModalComp_plateA}. The "signed-amplitudes" along one diagonal are represented in Fig.~\ref{fig:diag2_1}.
\ 
\subsection{Partial modal analysis of a rectangular plate (B): mid-frequency and moderate modal overlap ($30\%\leq\mu\leq50\%$).} 
\begin{figure}[ht!]
\centering
\includegraphics[width=0.9\linewidth]{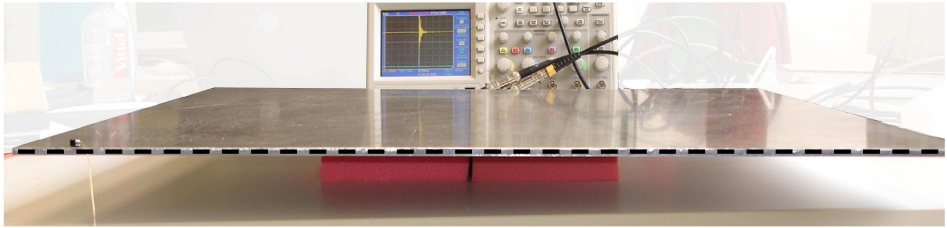}
\caption{Plate (B) with the line $x=0$ where modal analysis is performed.} 
\label{fig:plaquerectamort}
\end{figure}

The plate (\emph{AU4G}, $590 \times 637 \times 5.2$ mm$^3$) shown in Fig.~\ref{fig:plaquerectamort} is supported by four blocks of foam around the centre in order to ensure high damping; boundary conditions can still be considered as "free-free".
\ 

\begin{figure}[ht!]
\centering
\includegraphics[width=0.9\linewidth]{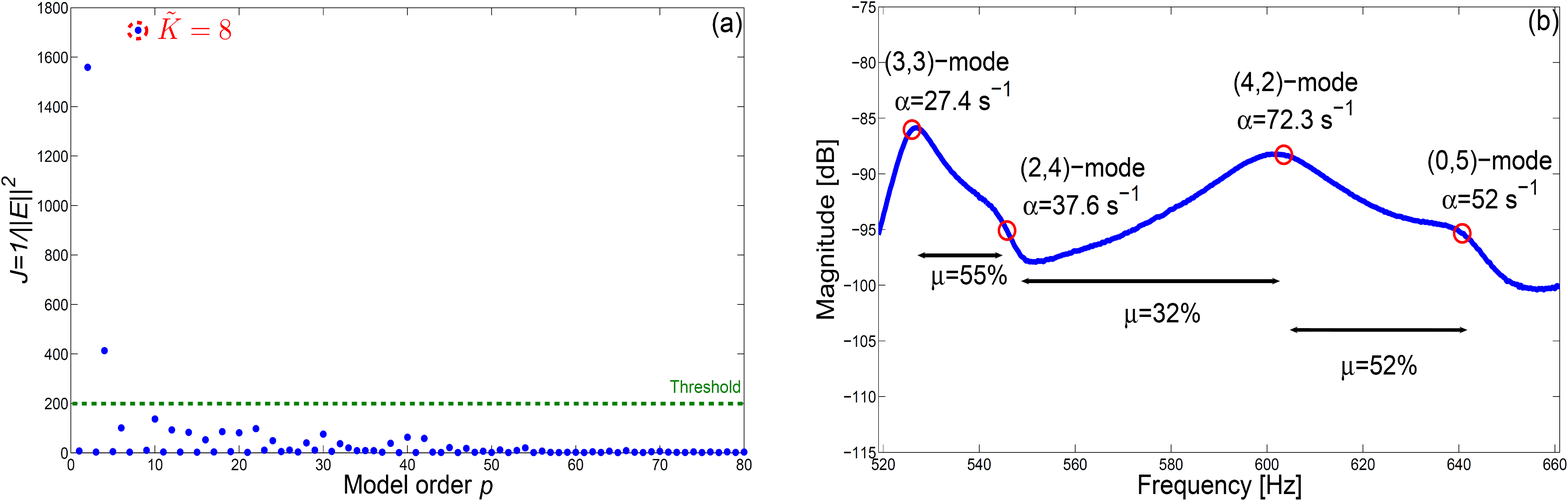}
\caption[Analysis in the frequency-band 520-660 Hz along one side of plate (B)]{Plate (B): modal analysis along the long side between 520 and 660~Hz. (a)~ESTER criterion on one response signal (point 32): the value ${\tilde{K}}=8$ (four modes) is detected; (b)~Amplitude of the Fourier spectrum at the same point. The length of the "useful" signal is $\approx 1.7$~s (it is masked by noise afterwards) and increasing amounts of zero-padding were tried: beyond a total length of 17~s for  the analysis window (as retained here), the spectrum does not change appreciably. \textcolor[rgb]{1,0,0}{$\circ$}~marks: modes estimated by ESPRIT.} 
\label{fig:analysis_plateB}
\end{figure}

The measurements are made at $33$ regularly spaced points along the long side ($x=0$). The sampling frequency is 50~kHz. The considered frequency-band is 520-660~Hz; the modal overlap is about $40\%$. In this mid-frequency region, a typical Fourier spectrum (Fig.~\ref{fig:analysis_plateB}(b)) does not exhibit well-separated modes. The result of the ESTER procedure is shown in Fig.~\ref{fig:analysis_plateB}(a), revealing four modes in this frequency-band.
\ 

\begin{figure}[ht!]
	\centering
	\includegraphics[width=1\linewidth]{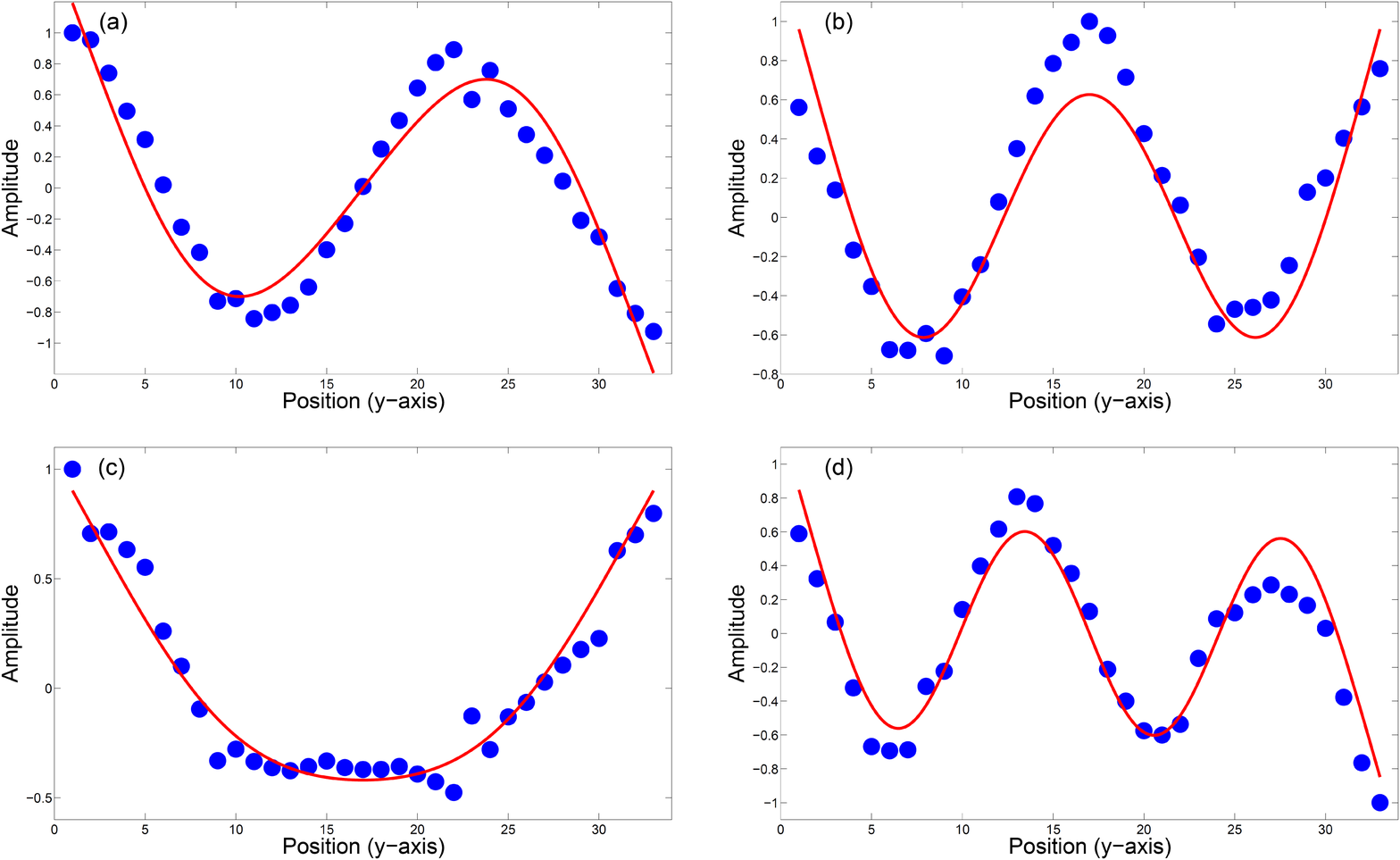}
	\caption[Modal analysis of plate (B) along one side]{Plate (B): modal analysis along one side (normalised "signed" amplitudes). \textcolor[rgb]{0,0,1}{$\bullet$}~ marks: \emph{measured} modes. \textcolor[rgb]{1,0,0}{Solid}~lines: \emph{theoretical} conservative modes. (a)~(3,3)-mode; (b)~(2,4)-mode; (c)~(4,2)-mode; (d)~(0,5)-mode.}
\label{fig:plaqueB_modalanalysis} 
\end{figure}

The modal shapes are represented by the "signed amplitudes" in Fig.~\ref{fig:plaqueB_modalanalysis}. Mass loading creates no visible asymmetry in modal shapes of plate B (its mass is 1.4$\cdot$10$^4$ times that of the accelerometer) and the negative shift of modal frequencies is about 0.1~Hz. With help of the theoretical analysis (three-term Rayleigh method), the measured modes can be identified as the (3,3)-, the (2,4)-, the (4,2)-, and (0,5)-modes, ranking 17 to 20 in the mode series. The estimations of the modal parameters are given in Table~\ref{tab:ModalComp_plateB} together with the corresponding approximate theoretical modal frequencies for Young's modulus E=7.4$\cdot$10$^{10}$~Pa, density $\rho$=2790~kg~m$^{-3}$, and Poisson's ratio $\nu$=0.33 as given by the manufacturer. A detailed discussion on the determination of theoretical modal frequencies, their dependency on material properties and plate geometry, and their comparison with experimental values is presented in the next section for plate (C).
\ 

\begin{table}[ht!]
\centering
\begin{tabular}{l c c c c}
   \hline
   
   \hline
\ &(3,3)-mode&(2,4)-mode&(4,2)-mode&(0,5)-mode\\
	\hline
	$f_\text{RAY}$~(Hz) & 523.7 & 542.7& 587.2& 645.7\\
	\hline
$f$ (Hz)&526.0&545.8 &603.5& 640.7\\
$\alpha$ (s\textsuperscript{-1})&27.4 &37.6 & 72.3 &52.0\\
 \hline
 
 \hline
 \end{tabular}
\caption{Plate (B): estimations of the modal parameters between 520 and 660~Hz. Top line: conservative plate treated by the improved Rayleigh method. Bottom line: experimental.}
\label{tab:ModalComp_plateB}
\end{table}
\ 

\subsection{Partial modal analyses of a rectangular plate (C): mid-frequency and high modal overlap ($45\%\leq\mu\leq70\%$).} \label{sec:PlateC}
In order to perform modal analysis on high-order modes ($\approx200$) near the acoustical coincidence frequency, a larger plate was considered (\emph{AU4G}, $1000~\times~1619~\times~5$~mm$^3$). Modes are analysed on a $10\ \times\ 10$ mesh with a 1~cm grid-step. Modal frequencies and damping factors are determined as the weighted means of the 100 corresponding estimations.

\begin{figure}[ht!]
\centering
\includegraphics[width=0.9\linewidth]{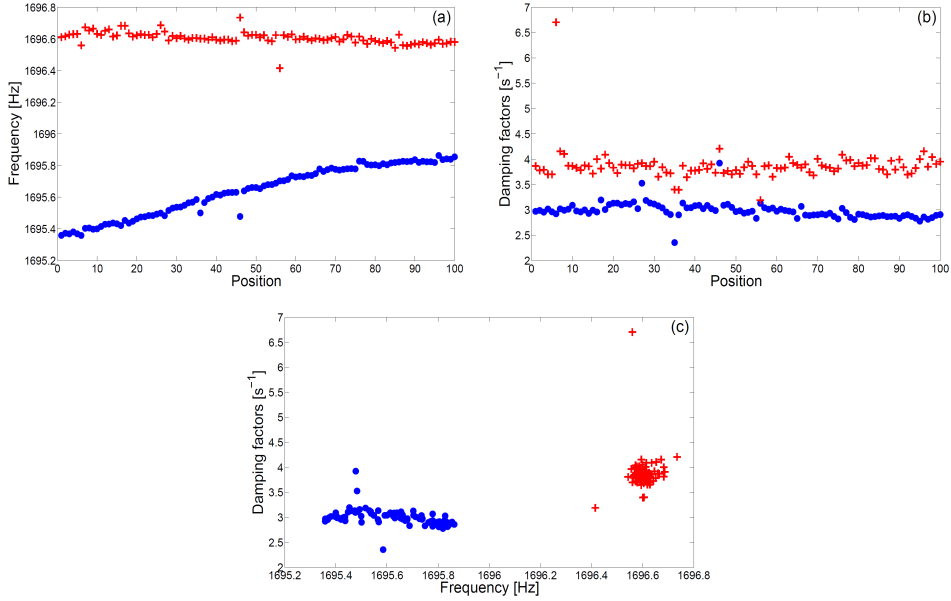}
\caption[Comparison]{Comparison of the two suspending setups for the (12,4)-mode of plate (C). \textcolor[rgb]{0,0,1}{$\bullet$}~marks: rubber-bands suspension.  \textcolor[rgb]{1,0,0}{+}~marks: nylon-lines suspension. (a)~Modal frequencies measured chronologically; (b)~Damping factors measured chronologically; (c)~Scattering of modal frequencies and damping factors.}
\label{fig:compar_1696}
\end{figure}

Two experimental setups were developed in order to ensure free-free boundary conditions for this 22.5~kg plate: suspension by six thick rubber bands glued along one side of the plate and suspension by two nylon lines passing trough small holes near the top plate edge. Both are presented in order to illustrate the sensitivity of the method. The experimental values of the modal frequencies are estimated with an accuracy of $\approx$~0.1~Hz (see Fig.~\ref{fig:compar_1696} and Table~\ref{tab:uncert}) in two narrow frequency-bands (around 1700~Hz and 2100~Hz). Theoretical values are determined as follows.

In the frame of the Kirchhoff-Love plate theory \cite{LOV1927}, the modal angular frequencies $\omega_{m,n}$ are given by: 
\begin{equation}\label{eq:modfreqlove}
\omega_{m,n}^2=B\,k_{m,n}^4
\end{equation}
where $B=\cfrac{E\,h^2}{12\,\rho\,(1-\nu^2)}\,=\cfrac{D}{\rho\,h}$ and $h$ is the thickness of the plate. The wavenumbers $k_{m,n}$ are determined by the plate dimensions $a$ and $b$ and by the boundary conditions. 

Since the physical parameters of the plate are not readily available with the desirable precision, we have estimated the $\,B(E,\rho,\nu,h)$ factor by comparing the 18\textsuperscript{th} to 25\textsuperscript{th} measured modal frequencies to those given by finite element simulations\footnote{Simulations are carried out with 8-node thin-shell elements (as in~\cite{HUR2007}). A mesh of 70~$\times$~100 elements is used, corresponding to $\approx$~35 points per wavelength at 200~Hz in the $x$-direction (respectively $\approx$~30 point in the $y$-direction).} (Table~\ref{tab:comparefreqFEM}). These particular modes are chosen because they are well-separated and the free-free boundary conditions are well-ensured. Minimising the average of the absolute values of the relative frequency differences between experiments and FEM simulations yields $B_{\text{FEM}}=61.0~\text{m}^4~\text{s}^{-2}$. With this estimated value\footnote{The values provided by the manufacturer for the duraluminium properties are E=7.4$\cdot$10$^{10}$~Pa, $\rho$=2790~kg~m$^{-3}$, and $\nu$=0.33. With these values, the value retained for $B_{\text{FEM}}$ corresponds to $h=4.96~\text{mm}$.}, the average relative difference in this frequency-band is $\approx 0.47\%$. Since the finite-element method introduces some spurious stiffness in the simulated system, the numerical value of $B_{\text{FEM}}$ is certainly slightly overestimated.

\begin{table}[ht!]
\centering
\begin{tabular}{c c c c c c c c c}
   \hline
   
   \hline
 	$f_\text{meas}$~(Hz) &125.9 & 139.3 & 141.1 & 147.5 & 150.6 & 154.7& 160.5 & 171.0\\
  $f_{\text{FEM}}$~(Hz) &125.5& 139.2&142.2&149.7 &151.2 &154.7 &159.5&170.9\\
 \hline
  $\cfrac{|f_\text{FEM}-f_\text{meas}|}{f_\text{meas}}\times 100$ &  0.30   &  0.12   &   0.80  &   1.49   &  0.38   &  0.01  &   0.62   &  0.08\\
 \hline
 
 \hline
 \end{tabular}
\caption{Plate (C): comparison between eight modal frequencies estimated by ESPRIT ($f_{\text{meas}}$) and calculated by a finite-element method ($f_\text{FEM}$).}
\label{tab:comparefreqFEM}
\end{table}
\

The modal frequencies and modal shapes of the high-order modes in the two frequency-bands of interest (around 1700 and 2100~Hz) are calculated with the approximate three-term Rayleigh method, using  the values estimated above for the physical parameters. According to Reference \cite{HUR2007}, the systematic error for the first modal frequencies calculated by this method is positive and less than +1\%.

\begin{figure}[ht!]
\centering
\includegraphics[width=1\linewidth]{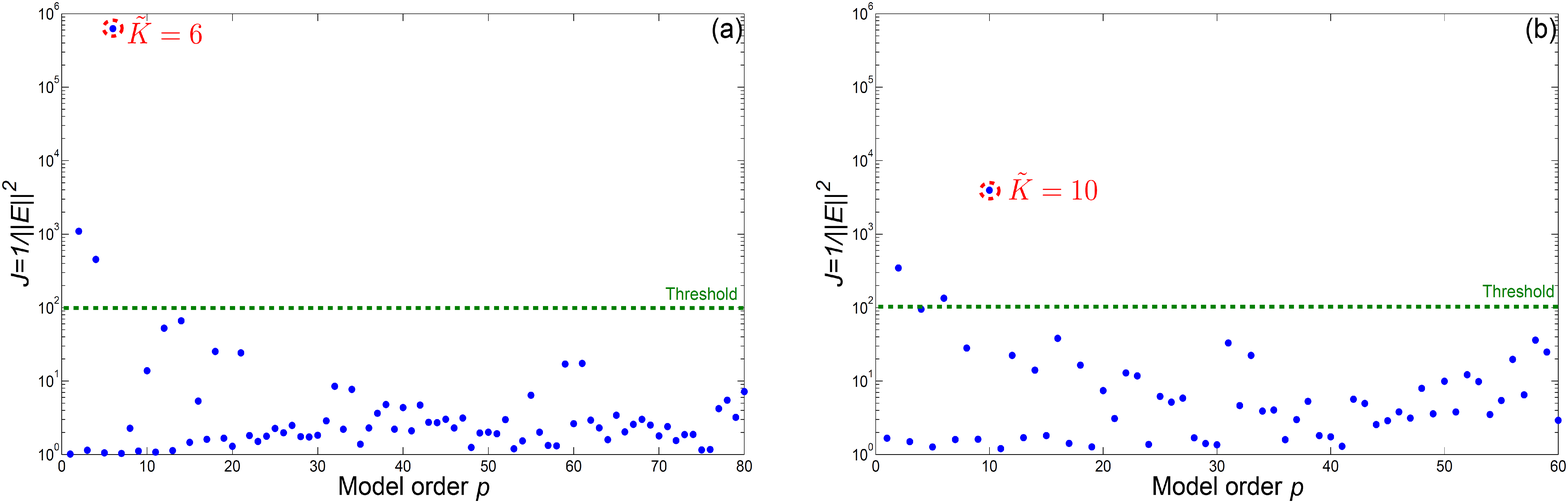}
\caption[ESTER criterion. Plate (C)]{ESTER criterion in two frequency-bands (Plate (C)). (a)~1685-1697~Hz frequency-band, point 1: the value ${\tilde{K}}=6$ (three modes) is detected. (b)~2065-2110~Hz frequency-band, point 11: the value ${\tilde{K}}=10$ (five modes) is detected.} 
\label{fig:ester_plateC}
\end{figure}

\begin{figure}[ht!]
\centering
\includegraphics[width=1\linewidth]{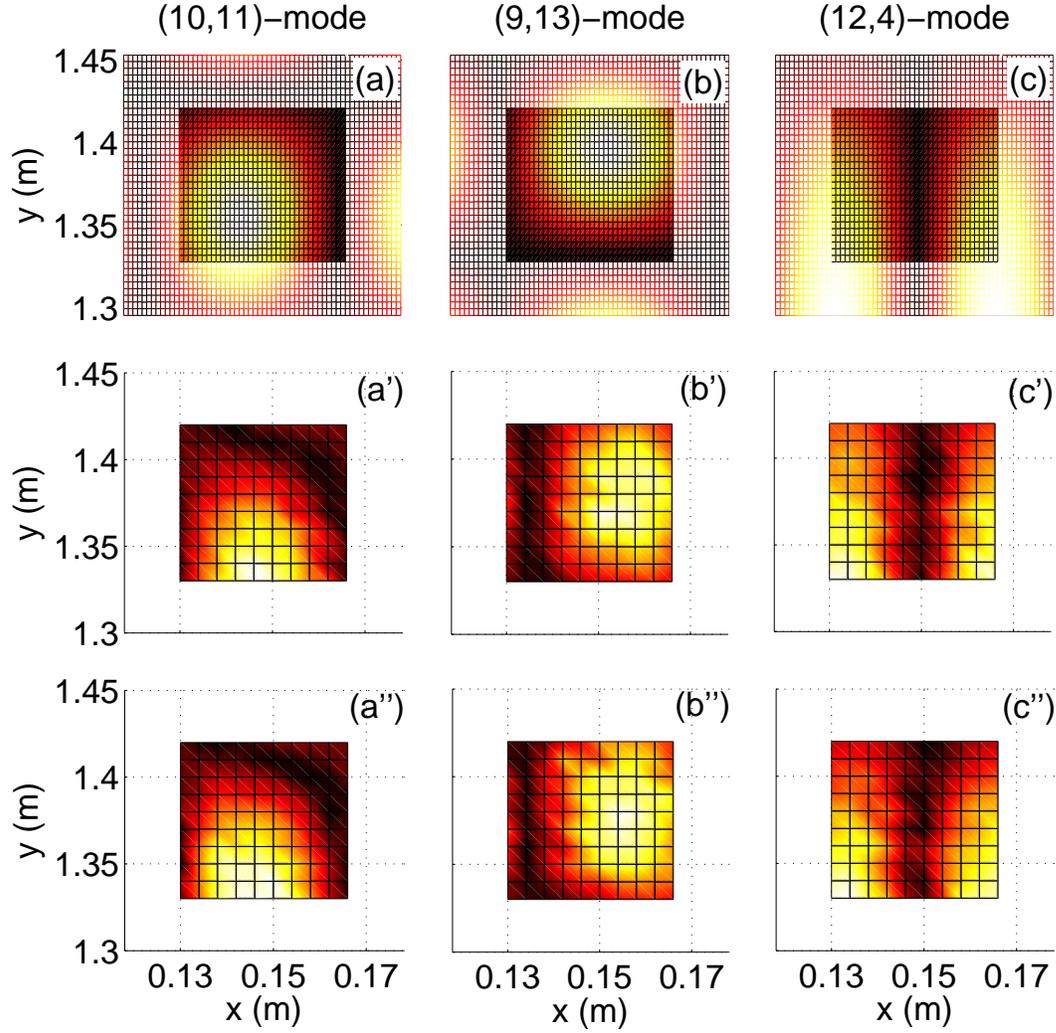}
\caption[Partial modal analysis of the plate (C)]{Plate (C): partial modal analysis between 1685 and 1697~Hz, with $\mu\simeq45\,\%$. (a),~(b),~(c)~Theoretical modal shapes obtained by the improved Rayleigh method. (a'),~(b'),~(c')~Measured modal shapes with the rubber-bands suspension. (a''),~(b''),~(c'')~Measured modal shapes with the nylon-lines suspension.}
\label{fig:plaqueC_1685_1697}
\end{figure}

The result of the ESTER procedure for the first frequency-band (1685-1697~Hz, $\mu\simeq45\%$) is shown in Fig.~\ref{fig:ester_plateC}(a) and the corresponding partial modal analysis results are given in Fig.~\ref{fig:plaqueC_1685_1697} and Table~\ref{tab:ModalComp}. Results for the three modes detected in this frequency-band are reported for both suspension arrangements. Also shown in Fig.~\ref{fig:plaqueC_1685_1697} and Table~\ref{tab:ModalComp} are the theoretical modal shapes and modal frequencies for modes (10,11), (9,13), and (12,4) which are the 199$^{\text{th}}$ to 201$^{\text{st}}$ modes. 
\ 
\ 
\begin{table}[ht!]
\centering
\begin{tabular}{l c c c c c c}
	\hline
	
	\hline
\ & \multicolumn{2}{c}{(10,11)-mode}& \multicolumn{2}{c}{(9,13)-mode}& \multicolumn{2}{c}{(12,4)-mode}\\
	\hline
	$f_\text{RAY}$~(Hz) & \multicolumn{2}{c}{ 1695.5 }& \multicolumn{2}{c}{1697.0}& \multicolumn{2}{c}{1703.4}\\
\hline
Suspension& rub.& nyl.&rub.& nyl.&rub.& nyl.\\
$f$ (Hz)&1689.9 &1690.0 &1693.1 &1693.3 &1695.7 &1696.6\\
$\alpha$ (s\textsuperscript{-1})&4.2 &4.6 & 4.3 &4.7 &3.0 &3.9\\
\hline

\hline
\end{tabular}
\caption{Estimations of modal parameters between 1685 and 1697~Hz for two suspension conditions. Top line: conservative plate treated by the improved Rayleigh method. Bottom line: experimental.}
\label{tab:ModalComp}
\end{table}
\ 

Measured and calculated modes match closely. The positions of the nodal lines are correct for the three modes. The measured modal shapes are almost identical for the two experimental setups. These results, together with the estimation of uncertainties (see below) illustrate the precision and reproducibility of the method.

The values of the calculated modal frequencies (Table~\ref{tab:ModalComp}) are systematically slightly larger than the measured ones by 0.2-0.5\%. Since this is also the case in the 2100~Hz frequency-band (see below), there must be a systematic error for wich we propose the following explanations. (a) The value of $B_{FEM}$ used for the calculation of the modal frequencies is overestimated. (b) The improved Rayleigh method overstimates modal frequencies\footnote{According to \cite{WAR1954}: \emph{By the Rayleigh principle, if a suitable waveform $W$ is assumed, satisfying approximately the boundary conditions, the resulting frequency value is very near, but higher than, the true value, because the assumption of an incorrect waveform is equivalent to the introduction of constraints to the system.}}. (c) In Kirchhoff-Love plate theory, the rotary inertia and the shear effects are ignored; for the plate considered here, the correction given by the more precise Mindlin theory \cite{MIN1951} in the $\omega(k)$ curve is around $-0.5\%$ at 1700~Hz.

Uncertainties reported in Table~\ref{tab:uncert} for the two suspension conditions are evaluated according to Eq.~\eqref{eq:uncert}:
\begin{equation}\label{eq:uncert}
\cfrac{\Delta{f}}{f}=\cfrac{\sqrt{\cfrac1{N_b}\displaystyle\sum_{i}(f_i-\bar{f})^2}}{\bar{f}}
\end{equation}
with $\displaystyle\bar{f}$ the weighted mean of the estimated modal frequencies and $N_b$ the number of estimations ($100$ in our case). This uncertainty estimate is pessimistic since deviations are not weighted here\ldots

\begin{table}[ht!]
\centering
\begin{tabular}{l c c c c}
	\hline
	
	\hline
\ &	\ & (10,11)-mode &(9,13)-mode & (12,4)-mode\\
	\hline
	\multirow{2}{28mm}{Rubber-bands suspension}& $\frac{\Delta{f\idr{rub}}}{f\idr{rub}}$ & $7.1\cdot10^{-5}$ & $9.0\cdot10^{-5}$ & $9.4\cdot10^{-5}$ \\ 
	\ & $\frac{\Delta{\alpha\idr{rub}}}{\alpha\idr{rub}}$ & $8.5\cdot10^{-2}$ & $6.7\cdot10^{-2}$  &  $5.3\cdot10^{-2}$\\ 
	\hline
	\multirow{2}{28mm}{Nylon-lines suspension}& $\frac{\Delta{f\idr{nyl}}}{f\idr{nyl}}$ & $4.0\cdot10^{-5}$ & $5.1\cdot10^{-5}$ & $2.2\cdot10^{-5}$ \\ 
	\ & $\frac{\Delta{\alpha\idr{nyl}}}{\alpha\idr{nyl}}$ & $8.5\cdot10^{-2}$ & $4.8\cdot10^{-2}$  &  $3.9\cdot10^{-2}$\\
	\hline
	
	\hline
  \end{tabular}
\caption{Uncertainties on frequencies and damping factors for the three modes between 1685 and 1697~Hz under two suspension conditions.}
\label{tab:uncert}
\end{table}
\  

During measurements with the first experimental setup, we noticed a drift in the estimation of the frequencies and possibly also in the estimation of the damping factors (see Fig.~\ref{fig:compar_1696}(a) for the chronological representation of these estimations). The second suspension setup (nylon lines in small holes) appears to be more stable (Figs.~\ref{fig:compar_1696}).
The overall +0.4~Hz frequency-drift  in the rubber-band case is larger than the uncertainty in the estimation of the modal frequency. The interpretation for the sign of the drift on frequency, for the fact that $f_\text{rub}<f_\text{nyl}$, and for similar observations on the damping factors goes as follows.
\ 
Rubber bands add a mass to the system. However, rubber bands slip slightly and the added mass decreases in time. This is also consistent with the very slight negative drift in the damping factor $\alpha_\text{rub}$. The $\alpha_\text{nyl}>\alpha_\text{rub}$ observation is interpreted by the fact that the vibrations of the plate are more strongly transmitted to the suspension frame by the nylon lines than by rubber bands.

\begin{figure}[ht!]\centering
\includegraphics[width=0.6\linewidth]{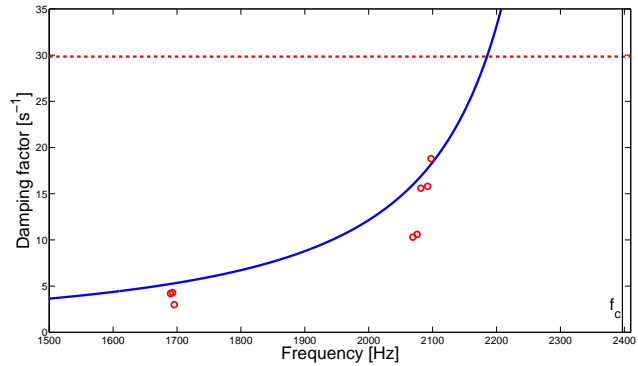}
\caption[Plate radiation]{Damping factors due to the radiation of an aluminium plate. Boundary conditions and radiation conditions are not the same for the experimental and for the theoretical determinations. {\color[rgb]{1,0,0}$\circ$}~marks: measured damping factors in the two frequency-bands where modal analysis was performed. \textcolor[rgb]{0,0,1}{Solid}~line: damping curve due to acoustical radiation of a baffled simply-supported plate (after~\cite{MAI1962}). \textcolor[rgb]{1,0,0}{Dashed}~line: (asymptotic) damping factor $\alpha$ of an infinite plate above the coincidence frequency $f\idr{c}$ ($\alpha_{\infty}=\rho\idr{a}c\idr{a}/(\rho h)$, where $\rho\idr{a}=1.2$~kg\,m$^{-3}$ is the density of air).}
\label{fig:radiate}
\end{figure}

In the second frequency-band (2065-2110~Hz) where modal analysis was performed, the modal overlap factor is $\approx 70\%$. Compared with the 1685-1697~Hz frequency-band, the important increase in damping factor (from $\approx4$ s$^{-1}$ to $\approx15$ s$^{-1}$) and thus in modal overlap is due to the sudden increase in acoustical radiation when the frequency approaches the coincidence frequency $f_\text{c}$. For this isotropic plate, $f_\text{c}$ is given (see \cite{REN1997} for example) by: 
\begin{equation}\label{eq:freqcrit}
f\idr{c}=\cfrac{c\idr{a}^2}{2\pi h}~\sqrt{\cfrac{12\rho(1-\nu^2)}{E}}
\end{equation}
where $c\idr{a}$ is the speed of sound in air ($\approx343\text{~m\,s}^{-1}$). Above this frequency, the wavelength of flexural waves in the plate is larger than the wavelength of acoustical waves in air and an infinite plate radiates sound; for a finite plate, the increase in radiation efficiency is gradual when the frequency approaches $f\idr{c}$ (see Fig.~\ref{fig:radiate}). In our case, the coincidence frequency is about 2.4~kHz.

\begin{figure}[ht!]
\centering
\includegraphics[width=1\linewidth]{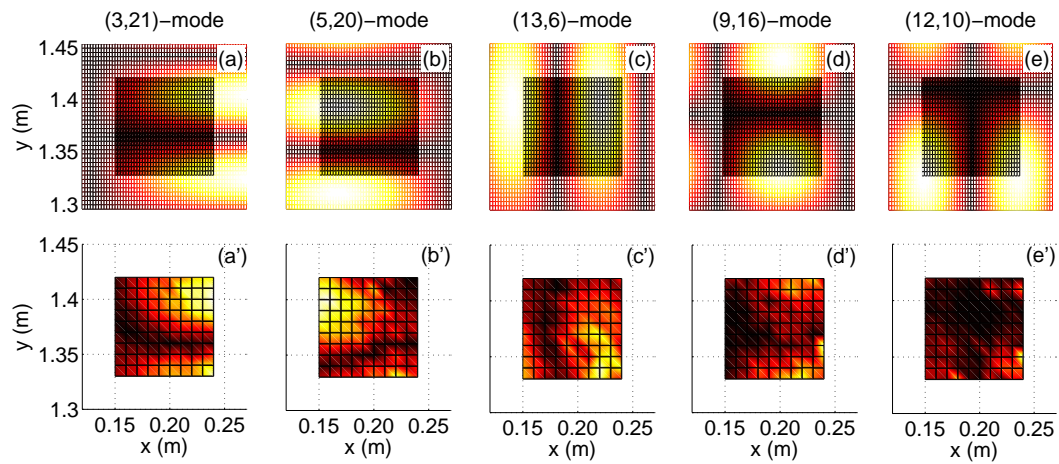}
\caption[Partial modal analysis of the plate (C)]{Plate (C): partial modal analysis between 2065 and 2110~Hz, with $\mu\simeq70\%$. (a),~(b),~(c),~(d),~(e)~Theoretical modal shapes obtained by the improved Rayleigh method. (a'),~(b'),~(c'),~(d'),~(e')~Measured modal shapes.}
\label{fig:plaqueC_2065_2110}
\end{figure}

Despite the high modal overlap factor, the ESTER procedure clearly detects the correct number of modes (Fig.\ref{fig:ester_plateC}(b)). The modal analysis results are given in Fig.~\ref{fig:plaqueC_2065_2110} and Table~\ref{tab:ModalComp_2} together with the results of calculations for the 243$^{\text{rd}}$ to 247$^{\text{th}}$ modes, corresponding to the (3,21), (5,20), (13,6), (9,16), and (12,10) modal shapes.
\ 

\begin{table}[ht!]
\centering
\begin{tabular}{l c c c c c c}
	\hline
	
	\hline
\ &(3,21)-mode&(5,20)-mode&(13,6)-mode&(9,16)-mode&(12,10)-mode\\
	\hline
	$f_\text{RAY}$~(Hz) & 2077.3 & 2086.6& 2096.7& 2100.3& 2112.6\\
	\hline
$f$ (Hz)&2069.1 &2075.7 &2081.5 & 2092.1 & 2097.5\\
$\alpha$ (s\textsuperscript{-1})&10.3 &10.6 & 15.6 &15.8 &18.8\\
\hline

\hline
\end{tabular}
\caption{Estimations of modal parameters between 2065 and 2110~Hz for the rubber band suspension. Top line: conservative plate treated by the improved Rayleigh method. Bottom line: experimental.}
\label{tab:ModalComp_2}
\end{table}
\ 

Matching is excellent for frequency values and correct for modal shapes. The higher values of calculated modal frequencies (Table~\ref{tab:ModalComp_2}) can be explained as in the 1700~Hz frequency-band. In the (12-10)-mode case, the analysed region is essentially nodal; the signal to noise ratio is $\approx\,$30~dB and the method clearly meets its limits.

Experimental results for the damping factors are displayed as a function of frequency in Fig.~\ref{fig:radiate}. Theoretical results \emph{in a different configuration} are available for the sake of an approximate comparison: in the case of simply supported baffled plate and under the assumption of the diffuse wavefield, Maidanik \cite{MAI1962} gives an analytical expression of the average damping due to radiation. The other contribution to damping of an aluminium plate is due to thermoelastic losses \cite{CHA2001}. The damping model established by Chaigne \emph{et al.} \cite{CHA2001} gives $\alpha_\text{therm}<0.14~\text{s}^{-1}$ for this aluminium plate. This value is very small compared with radiation damping in this frequency range of interest. It has therefore not been taken into account by the solid-curve in Fig.~\ref{fig:radiate}. The main physical difference between experimental and theoretical conditions lies in the acoustical short-circuit between the front and the back of the plate, reducing radiation efficiency and decreasing damping factors. This is compatible with the discrepancy between the measured points and the curve given by Maidanik.

\section{Conclusion}
The modal analysis method presented in this article resolves cases in which the Fourier transform meets its limits. Partial modal analyses of vibrating plates with high modal overlap (up to 70\%) match theoretical modal predictions. This method may contribute to filling the gap between the low-frequency and the high-frequency domains where Fourier modal analysis and statistical methods respectively apply. The ESTER technique appears as a good tool for estimating the modal density, an essential parameter for the study of vibrating structures in the mid- or high-frequency domains.

At frequencies larger than those presented here the results were not as satisfactory; this is mainly due to the signal-to-noise ratio limitation of the signal processing method. Moreover, the \emph{spatial resolution} of the method becomes also a limiting factor: the uncertainty in the position of the impact-excitation ($\approx 0.5$~cm) becomes significant compared with the grid-step (1~cm). However, a larger grid-step would not be acceptable at the considered wavelength (about 15~cm at 2.1~kHz for plate (C). 

The SNR limitation can be partly overcome by the use of a continuous excitation with a signal that allows the impulse response reconstruction by deconvolution techniques (\emph{swept-sine technique} as in  \cite{FAR2007}, for example).

\setcounter{section}{0}
\renewcommand{\thesection}{\appendixname \ \Alph{section}}
\renewcommand{\theequation}{\Alph{section}.\arabic{equation}}

\section{The three-term Rayleigh method}\setcounter{equation}{0}
\label{sec:Appendix_beam}
According to classical plate theory (see \emph{e.g.} \cite{TIM1974}), the maximum strain energy, or potential energy of bending $V$, of an isotropic rectangular thin plate is given by:
\begin{equation}\label{eq:V_max}
V_{max}=\cfrac1{2}\,D \int_0^a{\int_0^b{\,
\left[
\left(\cfrac{\partial^2W}{\partial x^2}\right)^2+
\left(\cfrac{\partial^2W}{\partial y^2}\right)^2+
2\nu\,\cfrac{\partial^2W}{\partial x^2}\,\cfrac{\partial^2W}{\partial y^2}+
2(1-\nu)\left(\cfrac{\partial^2W}{\partial x\partial y}\right)^2
\right]\,}}\text{d}y\,\text{d}x
\end{equation}
where $W$ is the modal shape and $D$ is $\cfrac{Eh^3}{12(1-\nu^2)}$. The maximum kinetic energy $T$ of the plate is: 
\begin{equation}\label{eq:T_max}
T\idr{max}=\cfrac{\rho\,h\,\omega^2}{2} \int_0^a \int_0^b\,
W^2\text{d}y\,\text{d}x
\end{equation} 

The Rayleigh principle yields the modal angular frequency $\omega$:
\begin{equation}\label{eq:omega_Ray}
\omega^2=\cfrac{2\,V\idr{max}}{\rho\,h\, \displaystyle\int_0^a \int_0^b\,
W^2\text{d}y\,\text{d}x}
\end{equation}

Kim \& Dickinson \cite{KIM1985} extend the Rayleigh method \cite{WAR1954} by considering three terms in $W$: 
\begin{equation}\label{eq:rayleigh_imp}
W_{(m,n)}(x,y)={X_m(x)Y_n(y)-c\,X_{m'}(x)Y_n(y) -d\,X_m(x)Y_{n'}(y)}
\end{equation}
where $X_m(x)$ (respectively $Y_n(y)$) is the $m$ (resp. $n$)-th order normal modal shape of a beam with the same boundary conditions as the corresponding edges of the plate; $X_{m'}(x)$, $Y_{n'}(y)$ are the next higher beam modal shapes, and $c$ and $d$ are constant quantities given below. In our case, boundary conditions are free-free: $m'=m+2$ and $n'=n+2$. The modal deflection $X_m$ of a free-free beam are given in Eq.~(11) of reference \cite{WAR1954}.

By substituting Eq.~\eqref{eq:rayleigh_imp} into Eq.~\eqref{eq:omega_Ray} and~\eqref{eq:V_max}, the modal angular frequency is:
\begin{equation}\label{eq:freqparam}
\omega_{m,n}^2=\cfrac{D\,\pi^4}{\rho\,h\,a^2b^2}\,\ \cfrac{C_{m,n}+c^2C_{m,n+2}+d^2C_{m+2,n}-2cE_{m,n}-2dE_{n,m}+2cdF}{1+c^2+d^2}
\end{equation}
with $$
\begin{array}{l}
C_{m,n}=\,G_m^4\cfrac{b^2}{a^2}+G_n^4\cfrac{a^2}{b^2}+2(\nu H_m H_n+(1-\nu)J_mJ_n)\\
E_{m,n}=\,\nu H_m(K_n+L_n)+2(1-\nu)J_mM_n\\
F_{m,n}=\,-\nu(K_mK_n+L_mL_n)+2(1-\nu)M_mM_n
\end{array}
$$
The values of $G_m$, $H_m$, $J_m$, $K_m$, $L_m$, $M_m$ are given in references \cite{WAR1954}~(Table~1) and \cite{KIM1985}~(Table~1). Finally, $c$ and $d$ can be determined by using the minimum potential energy theorem: 
$$
\begin{array}{c}
\cfrac{\partial V_{max}}{\partial c}=\cfrac{\partial V_{max}}{\partial d}=0\quad\Rightarrow\quad
\left\{\begin{array}{cc}
c=\cfrac{C_{m+2,n}E_{m,n}-E_{n,m}F}{C_{m,n+2}C_{m+2,n}-F^2}\\
d=\cfrac{C_{m,n+2}E_{n,m}-E_{m,n}F}{C_{m,n+2}C_{m+2,n}-F^2}
\end{array}
\right.
\end{array}
$$

\bibliographystyle{elsart-num}
\bibliography{ArticleV5}

\end{document}